\documentclass[useAMS,usenatbib]{mn2e}

\usepackage{amsmath}

\usepackage{mathtools}

\usepackage{graphicx}

\usepackage{txfonts}

\title[Models of solar surface dynamics: impact on eigenfrequencies and radius]{Models of solar surface dynamics: impact on eigenfrequencies and radius} 
\author[L. Piau et al.]{L. Piau$^{1}$\thanks{E-mail:
lrnt p@yahoo.com}, R. Collet$^{2}$, R. F. Stein$^{3}$, R. Trampedach$^{4}$, P. Morel$^{5}$ and S. Turck-Chieze$^{6}$\\
$^{1}$52, Avenue d Italie, 75013, Paris, France\\
$^{2}$Research School of Astronomy and Astrophysics, Australian National University, Canberra ACT 2611, Australia\\
$^{3}$Michigan State University, Department of Physics \& Astronomy East Lansing, MI 48824-2320, USA\\
$^{4}$JILA, University of Colorado and National Institute of Standards and Technology, 440 UCB, Boulder, CO 80309, USA\\
$^{5}$Observatoire de la Cote d'Azur, Boulevard de L'Observatoire, BP 4229 F06304 Nice Cedex 4, France\\
$^{6}$Service d'astrophysique, CEA Saclay, 91191, Gif-sur-Yvette, France}
\begin{document}



\date{Accepted to MNRAS}

\pagerange{\pageref{firstpage}--\pageref{lastpage}} \pubyear{2013}

\maketitle

\label{firstpage}

\begin{abstract}

We study the effects of different descriptions of the solar
surface convection on the eigenfrequencies of p-modes. 
1-D evolution calculations of the whole Sun and 3-D hydrodynamic
and magnetohydrodynamic simulations of the current surface are performed. 
These calculations rely on realistic physics.
Averaged stratifications of the 3-D simulations are introduced 
in the 1-D solar evolution or in the structure models.
The eigenfrequencies obtained are compared to those 
of 1-D models relying on the usual phenomenologies of convection and
to observations of the MDI instrument aboard SoHO.
We also investigate how the magnetic activity could change 
the eigenfrequencies and the solar radius, assuming that, 
3 Mm below the surface, the upgoing plasma advects 
a 1.2 kG horizontal field.

All models and observed eigenfrequencies are fairly close
below 3 mHz. Above 3 mHz the eigenfrequencies of the 
phenomenological convection models are {\it above} the observed 
eigenfrequencies. The frequencies of the models based 
on the 3-D simulations are slightly {\it below} the
observed frequencies. Their maximum deviation is
$\approx$ 3 $\mu$Hz at 3 mHz but drops below
1 $\mu$Hz at 4 mHz. Replacing the hydrodynamic 
by the magnetohydrodynamic simulation increases the eigenfrequencies. 
The shift is negligible below 2.2 mHz and then increases linearly with
frequency to reach $\approx$ 1.7 $\mu$Hz at 4 mHz. The impact of
the simulated activity is a 14 milliarcsecond shrinking of the solar layers 
near the optical depth unity.

\end{abstract}



\section{Introduction}\label{sec1}

Modelling the few megameters below the surface of 
solar-type stars requires to take account 
of a very rich and complicated physics 
(\cite{nsa09}). The stellar medium goes 
from optically thick and fully ionized 
to optically thin and neutral. This necessitates 
a sophisticated equation of state and forbids 
the use of the diffusion approximation for photons as
in stellar interiors. The wavelength
dependent radiative transfer equation must be 
addressed directly instead (\cite{nord82}).
Large scale physical processes also contribute
to the complexity of the region : the radial and latitudinal 
rotation profiles change right below the surface 
(\cite{basu99}; \cite{corbard02}),
the convection is extremely turbulent and
becomes sonic near the visible surface, 
thus compressibility effects and turbulent pressure 
cannot be neglected as deeper into the interior.
Finally, the magnetic activity strongly
affects those regions.

Both helio- and asteroseismological observations have 
pointed out the weaknesses in the understanding 
of the surface. In the Sun's case, the absolute 
frequencies that are observed are systematically below those
that are computed from models 
(\cite{TC97}; \cite{rabello08}; but see \cite{rosenthal99}). The difference has 
been identified as related to surface effects and
consequently dubbed the surface term. Addressing the surface term in the
solar case is not only interesting with respect to
solar physics but is also important for other low mass 
stars where mode identification is more difficult than in the Sun. 

Following the work of \cite{rosenthal99} (hereafter R99) now 
with a discussion of magnetic effects, we intend to compare various
solar models and their absolute p-mode frequencies.
These 1-D models are computed with the same 1-D stellar
evolution code. They only differ with respect
to the treatment of surface convection and/or surface
activity. Two of them are buildt using traditional
phenomenologies: the mixing length theory (\cite{bohm58}) 
and the more sophisticated full spectrum of turbulence model 
(\cite{cgm96}). Numerical simulation have 
outlined the limits of the phenomenological 
models in addressing the surface convection (\cite{ludwig02};
\cite{ludwig06})
The other models rely on 1-D horizontal and time 
averages of 3-D simulations of surface
convection. Some of these models use purely 
hydrodynamic 3-D simulations while others 
rely on magnetohydrodynamic 3-D simulations.

In the following we first describe how the 1-D solar
models are buildt (\S \ref{sec2}). We address the 3-D 
hydrodynamic surface simulations (\S \ref{sec21}), 
give the main inputs to our secular evolution code 
(\S \ref{sec22}) and describe the calibration process (\S \ref{sec23}).
In \S \ref{sec3} we compare the absolute frequencies of our models to the 
observed ones. We first focus on the hydrodynamic effects
considering the frequencies about the minimum of the activity 
cycle (\S \ref{sec31}). Then we address the impact of magnetic
field on the absolute frequencies and on the photospheric radius 
(\S \ref{sec32}).
We discuss our results and conclude in \S \ref{sec4}.

\section{Building the solar models}\label{sec2}

We treat the surface convection using 3-D (magneto)hydrodynamic 
simulations or the customary 1-D phenomenologies.
Both approaches provide us with the average vertical 
structure of the upper solar convection 
zone. For 3-D simulations, the mean stratification includes effects of the
interaction between radiative transfer and convective motions,
turbulent pressure and the magnetic 
field. These effects are directly or indirectly included as boundary
conditions to the 1-D solar structure. 
In order to estimate their impact, 1-D solar models are also buildt in the 
frameworks of the usual phenomenologies of surface convection: 
the mixing length theory and the full spectrum of turbulence phenomenology.
In the following, we first describe the 3-D simulations then turn 
to the 1-D solar secular models and to how 3-D effects are incorporated 
into them.

\subsection{Surface convection simulations}\label{sec21}

We use the Stagger magnetohydrodynamic code 
(\cite{stein98}; \cite{beeck12}). Stagger 
is a 'box in the star' type code. Our
simulation domain is 6000 $\times$ 6000 km horizontally and
extends from $\sim$940 km above the Rosseland optical depth $\rm \tau_{Ross}=1$ to 
$\sim$2900 km below. $\rm \tau_{Ross}=1$ corresponds to -69 km 
on the geometrical scale of the 3-D domain. The equations of compressible 
magnetohydrodynamics are solved explicitly for a few hours 
of solar surface time and over $240^{3}$ cells. The mesh resolution 
in both horizontal directions is constant at 25 km but varies in the vertical
direction to catch the rapid variation in the physical conditions.
Being as low as about 7 km near optical depth unity and the underlying 
superadiabatic region it is as high as 33 km at the lower boundary of the
domain. Performing solar surface hydrodynamic simulations
\cite{robinson03} have shown that the average
superadiabatic gradient and turbulent pressure (both important
quantities for the structure calculation of the upper convection zone)
are nearly independent of the horizontal extent of the box when varied
from 1350 km to 5400 km. This result was reached for
a 2800 km depth, comparable to ours, but fewer grid points
and much coarser cells sizes than in our calculations.
We are therefore confident that the current domain extension and mesh 
refinement are sufficient for the purpose of the calculations
(See also \cite{stein98}
for a discussion on resolution).


The gravity field is assumed constant at $\rm log g=4.437$ 
throughout the domain. The density
and the specific internal energy
of the plasma entering the domain from the lower edge 
are adjusted to $\rm 2.23\,10^{-5} g.cm^{-3}$ and 
$\rm 9.87\,10^{12} erg.g^{-1}$ respectively in order to 
obtain a time averaged effective temperature of 5772 K. 
At the solar surface the medium goes from optically thick
to optically thin and from almost fully ionized to almost 
completely neutral. One cannot avoid a sophisticated 
microphysics to describe these regions. 
We have adapted the OPAL 2005 equation of state
to Stagger with the solar surface composition prescription
of \cite{asplund09} (hereafter A09): the hydrogen 
and metal mass fraction are respectively X=0.7381 and Z=0.0134.
The contribution to heating and cooling due to radiation is accounted 
for by solving the radiative transfer equation at each time-step 
during the simulation along rays passing through all points at 
the optical surface. We use a Feautier-like (\cite{feau64}) 
long-characteristic solver and consider eight inclined directions
plus the vertical for the rays. 
We also account for non-grey effects by sorting out wavelengths into 
twelve opacity bins (\cite{nord82}; \cite{skart00}) according to 
their relative strength and to spectral region, then solving the radiative
transfer equation assuming an appropriate average opacity and a collective 
(integrated) source function for each bin. The monochromatic continuous 
and line opacities come from an updated version of the Uppsala opacity
package (Trampedach et al., in prep.; \cite{gusta75}; Plez, priv. comm.) 
and average bin opacities are computed according to the method described by 
\cite{collet11}. The monochromatic source functions are assumed to 
be purely thermal (i.e. equal to the Planck function). The opacities used to 
form the binning and solve the transfer are also based 
on the solar composition derived by A09.

We performed two solar convection simulations : a purely
hydrodynamic one and a magnetohydrodynamic one. 
Both models are started from a previously 
hydrodynamic simulation relying on the MHD equation of
state (hereafter EoS). 
Thus the changes in the thermal average structure or
any quantity related to the dynamics of the convection
stem from the differences between the MHD and the 
OPAL EoS (see \cite{trampedach06}
for a comparison of the two EoS). The simulations are first
run over 6000 s of solar surface time. Then 24
snapshots are stored every 5 minutes over
the next 2 hours of solar time. The temporal and spatial averages are computed
from these last snapshots. As we are interested in the average 
structure of the solar upper layers we have to ensure that 
these models are thermally and statistically relaxed. 

{\it i) Mass and energy fluxes}: 
While the mass fluxes at the upper boundary are 
negligible the average incoming mass flux at the lower boundary 
is $\rm 7.1\,10^{-1} g.cm^{-2}.s^{-1}$. 
Its difference to the average outgoing mass flux 
is about $\rm 4.4 10^{-4} g.cm^{-2}.s^{-1}$. Throughout 
the box the upgoing/downgoing mass flux difference is always less 
than $\rm 10^{-3} g.cm^{-2}.s^{-1}$ and there is no mass
redistribution along the simulation time.
At a given (snapshot) moment the lower boundary energy flux differs
by up to 44\% from the nearly constant upper 
radiative flux. However this difference shows no 
temporal trend and on average the incoming and
outgoing fluxes differ by less than 5\%.

\begin{figure}
\centering
\includegraphics[angle=90,width=8cm]{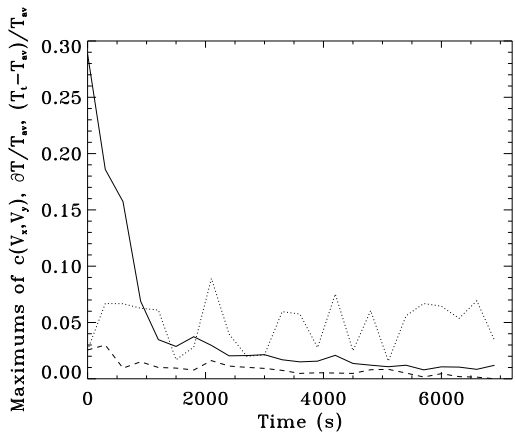}
\caption{Solid line: time evolution of the maximum 
of the absolute value of $\rm c(v_x,v_y)$,
the correlation between velocities in both horizontal directions. 
Dotted line: time evolution of the maximum 
of the absolute value of the temperature fluctuation with respect
to the average temperature. 
Dashed line : time evolution of the maximum 
of the absolute value of the difference between the 
temperature averaged until that time and the
temperature averaged over the whole simulation duration.  
\label{fig1}}
\end{figure}

{\it ii) Temperature and pressure profiles}:
Figure~\ref{fig1} shows the evolution of the maximum of the
fluctuations in temperature considered over the whole simulation box. 
Actually this maximum always occurs 
within 300 km around $\rm \tau_{Ross}=1$. 
It does not evolve with time. Figure~\ref{fig1} also shows that, at any depth,
the average horizontal temperature averaged over 1 hour or more remains 
within 2 percent of the average horizontal temperature 
over the whole simulation duration. 
This illustrates that the temperature averages become 
nearly independent of the integration time after 1 hour of solar
surface time.
The behavior of the pressure is comparable to that of the temperature.

{\it iii) Velocities}: the maximum velocity throughout
the box is remarkably constant around $\rm 14\,km.s^{-1}$. The maximums 
are always located in the superadiabatic layer between $\rm \tau_{Ross}=1$
and 250 km below. 
The correlation coefficient between velocities along
horizontal axes is $\rm c(v_x,v_y)=\frac{\overline{v_x
 v_y}-\overline{v_x}\,\,\,\overline{v_y}}{\sigma (v_x) \sigma (v_y)}$,
where the bars denote horizontal and time averages
and the $\rm \sigma$s denote the standard deviations e.g. : 
$\rm \sigma (v_x) =\sqrt{\overline{v_x^2}-\overline{v_x}^2}$. 
\cite{robinson03} simulations suggest $\rm c(v_x,v_y)$ is actually 
the slowest quantity to converge
statistically. Figure~\ref{fig1} show that it
falls steeply during the first 30 minutes.
It is below 0.03 at any depth after 40 minutes of simulation
and continues to decrease as time elapses.

The convergence of quantities such as temperature, pressure
or convective velocities is crucial to us. We aim at using
the averages of these quantities directly in the 1-D models
or to rely on them to compute the thermal
superadiabatic gradient and the turbulent pressure:
we use the average of temperature vs. gas pressure relation 
to derive the thermal gradient $\rm {dlnT}/{dlnP}$ 
\footnote{All the derivatives of this work are obtained 
thanks to the IDL deriv.pro routine that 
performs a numerical differentiation using a 3 points
Lagrangian interpolation.}.
In the following this gradient
is dubbed the hydrodynamic gradient $\rm \nabla_{HYD}$
or magnetohydrodynamic $\rm \nabla_{MHD}$
gradient owing to its calculation 
from the two different sets of 3-D simulations. Note that the turbulent pressure
is computed according to the approximate formula 
$\rm P_{turb}=\rho (\overline{v_z^2} - \overline{v_z}^2)$ where $\rm \rho$ and $\rm v_z$ 
respectively denote density and vertical
velocity. In the solar interior this approximation differs by less than
$\rm 15\%$ from the exact turbulent pressure.

\begin{figure}
\centering
\includegraphics[angle=90,width=8cm]{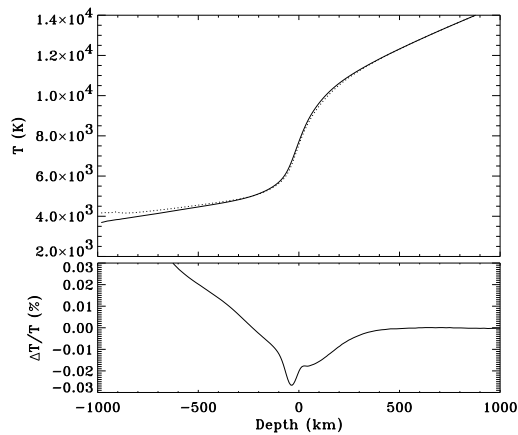}
\caption{Upper panel : average temperature profile vs depth 
of the hydrodynamic simulation (Solid line) and the
magnetohydrodynamic model (Dotted line). Lower panel : 
relative variation of the average temperature profile between 
the hydrodynamic simulation and the magnetohydrodynamic
simulation. \label{fig2}}  
\end{figure}

\begin{figure}
\centering
\includegraphics[angle=90,width=7cm,height=7cm]{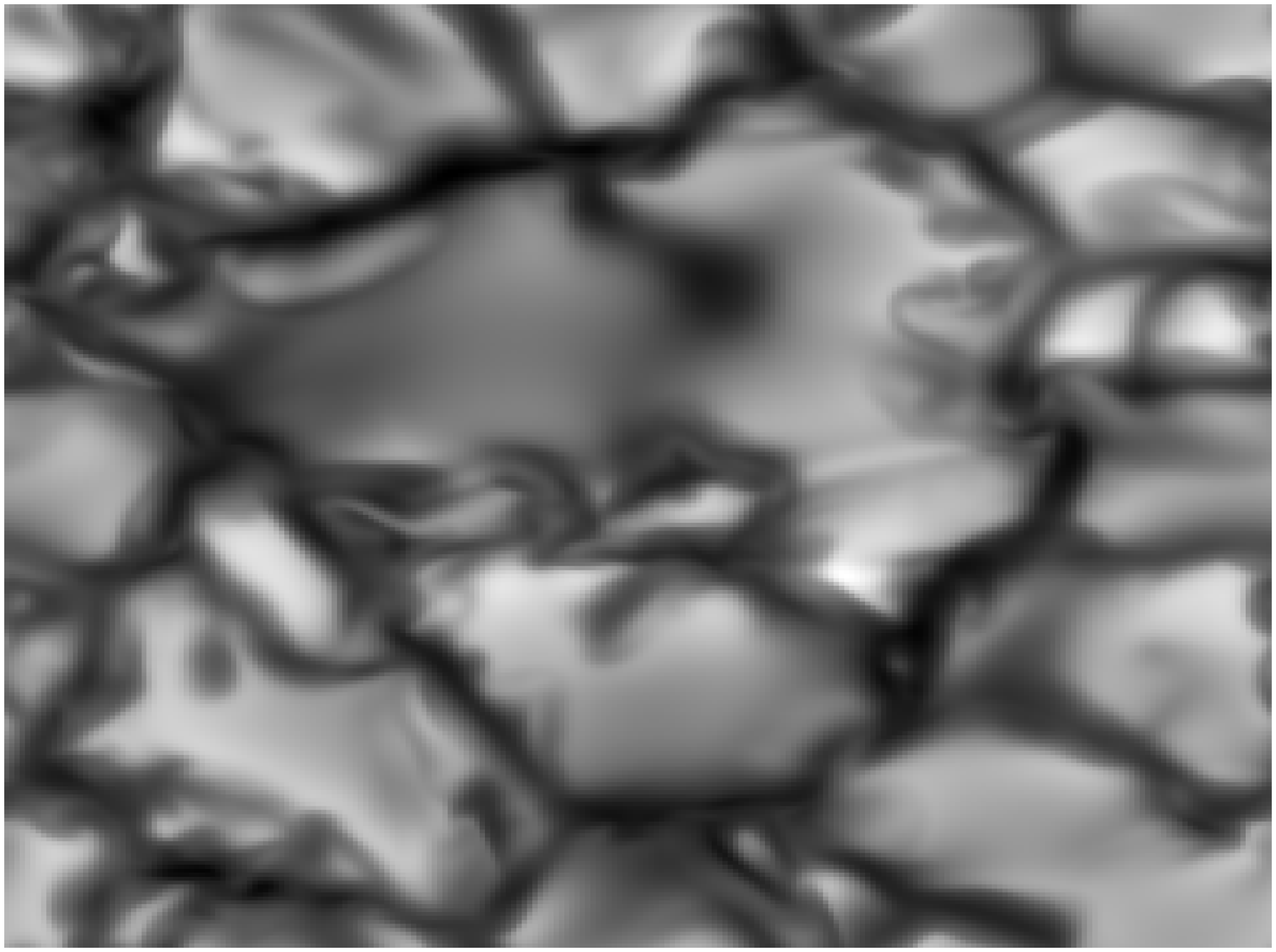}
\includegraphics[angle=90,width=7cm,height=7cm]{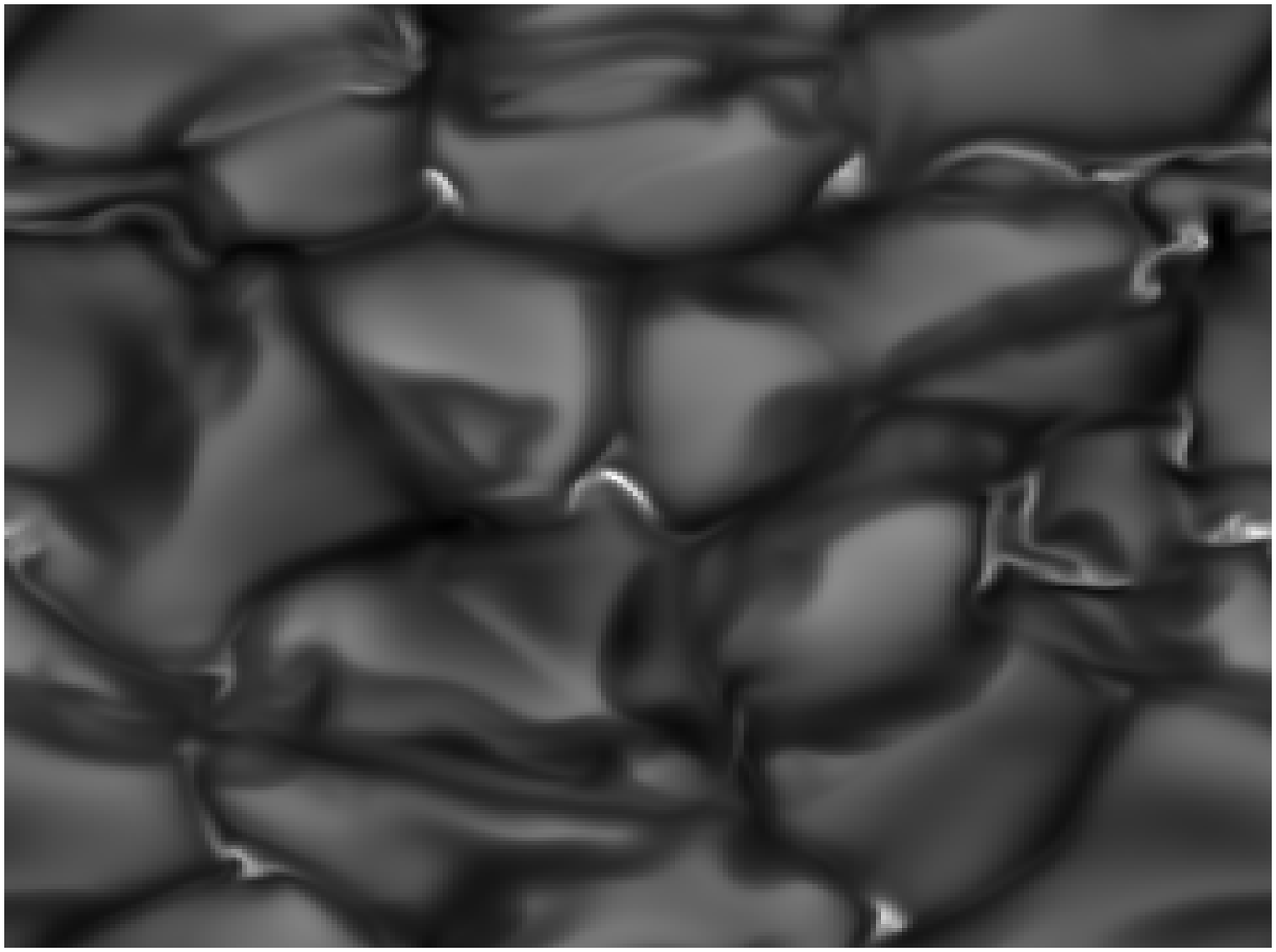}
\caption{Upper panel : continuum emerging intensity of the last
snapshot of the hydrodynamic simulation. The range is 0.65 to 1.4 the 
average emerging intensity. Lower panel :  continuum emerging intensity of the last
snapshot of the magnetohydrodynamic simulation. The range 
is 0.65 to 2.0 the average emerging intensity.
\label{fig3}}  
\end{figure}

In the magnetohydrodynamic simulation a 1.2 kG uniform, untwisted, 
horizontal field is advected into the computational domain by inflows 
at the lower boundary. In the outflows the vertical derivative of 
all the field components vanishes. The turnover time from 2.9 Mm is 
about 1 hour, so the calculation was relaxed for 1.6 turnover times 
before data started to be collected. The convective flows produce 
a hierarchy of serpentine loops with smaller ones riding piggy back 
on the larger one. Such simulations have produced bipolar flux emergence 
and magnetic field distributions that agree well with observations 
(\cite{stein12}). Besides the presence 
of a magnetic field, the magnetohydrodynamic simulation
relies on exactly the same physical inputs and parameters as the
purely hydrodynamic one and is started from the 
same initial state. Its 24 snapshots are used to estimate
the same quantities as in the 
hydrodynamic simulation plus the distribution
of the magnetic field. For optical depths larger than
$\rm \tau_{Ross}=10^{-1}$, 
the physical quantities described above 
are weakly affected by the magnetic field, the 
most significant changes occuring only within 
0.5 Mm of the visible surface. 
For instance the average temperature profile changes
by at most $\rm \sim 3\%$ at $\rm \tau_{Ross} \sim 4$ with respect
to purely hydrodynamic simulations (see Figure \ref{fig2}). The rms of turbulent 
velocities are an exception: they are decreased 
by $\rm \approx 10\%$ all the way down the simulation bottom.
From the energy density point of view the gas pressure
is always much larger than the turbulent pressure, the maximum
ratio being at $\rm 14\%$ in the superadiabatic region.
In turn the magnetic pressure only represents 
$\rm \approx 35\%$ of the turbulent pressure
throughout the superadiabatic region. We remark that
this ceases to be true deeper down or above the visible
surface where the magnetic pressure dominates a vanishing
turbulent pressure. Finally we mention that the ratio of
the rms of the horizontal magnetic field over $\rm \rho^{1/3}$
(with $\rm \rho$ the average density) does not depend on depth.
This is a characteristic feature 
of relaxed magnetohydrodynamic simulations.
At $\rm \tau_{Ross}=1 \, (\rho=2.1\, 10^{-7} g.cm^3)$ the rms of the 
horizontal and vertical magnetic fields are respectively $\rm B_h =240 \,G$
and $\rm B_v =140 \,G$. Figure \ref{fig3} displays the emerging continuum 
intensities from the last snapshots of the hydrodynamic
and magnetohydrodynamic runs. In agreement with simulations 
exhibiting similar surface fields (\cite{stein11}), the field intensity 
considered has no significant effect on the shape of the convective
cells. However bright points appear in the downflows of magnetohydrodynamic 
simulated surface and the maximum to minimum ratio of emerging 
intensity increases.

\subsection{Solar secular evolution}\label{sec22}

We use a modified version (\cite{piau11}) 
of the hydrostatic one-dimensional CESAM code (\cite{morel97}; \cite{morelebreton08}).
CESAM's calculations of the stellar interior rely on the usual 
ingredients of evolution codes:

{\bf The equation of state and opacities} are the OPAL2005 (\cite{rogers96}; \cite{rogers02}) ones 
for the \cite{asplund09} (A09) metal repartition.
Below $\rm logT = 3.75$ we use the \cite{ferguson05} 
opacities for the same metal repartition.

{\bf The composition changes}. The nuclear reaction rates are adapted from the 
NACRE compilation (\cite{angulo99}). The nuclear network is restricted to the
reactions relevant to the main sequence evolution : proton-proton 
chains and CNO cycles. In addition to the nuclear reactions the 
gravitational settling of elements is accounted for following
the prescriptions of \cite{proffitt93}.
It results in the usual abundance decrease in the photosphere 
helium and metals mass fractions from the zero age main 
sequence (hereafter ZAMS) to the actual age.

The outer convection and atmosphere are handled in a less traditional manner,
as we introduce prescriptions coming from Stagger 3-D calculations.
However for the purpose of comparison, models relying
on the usual 1-D convection prescriptions and atmospheres
are also computed.

{\bf Surface and 1-D convection phenomenologies}

The mixing length theory\footnote{ The detailed prescription of the MLT
we use is given in the appendix of \cite{piau05}, 
it is very similar to that of \cite{bohm58}.} 
(hereafter MLT) and the 
phenomenology of \cite{cgm96} (hereafter CGM) 
have been used for surface 
convection energy transfer and atmosphere calculation.
As stressed by \cite{montal04} it is necessary to use
the same phenomenology in the atmosphere and the interior.
Therefore we consider different atmosphere grids depending on the
underlying convection prescription. The approach is the 
same as in \cite{piau11} where the reader can find all 
the details of the implementation.

Both MLT and CGM atmosphere grid models were
computed with the 1-D Atlas12 code atmosphere structure code (\cite{castelli05}).
Atlas12 calculates the non-grey atmosphere stratification.
In both sets of atmosphere models, the current solar surface 
composition of A09 is considered and the convection characteristic 
length scale is $\rm 0.5 H_p$ (\cite{samadi06}). The atmosphere thermal gradient 
$\rm \nabla_a={dlnT}/{dlnP}$ is 
computed following the usual equation $\rm \nabla_a=\nabla_{rad} * \frac{df}{d\tau_{Ross}}$, 
$\rm \nabla_{rad}$ being the radiative diffusive gradient, and $\rm f(\tau_{Ross})$ the function 
relating effective temperature, optical depth and temperature:
$\rm T^4=\frac{3}{4}T_{eff}^4 f(\tau_{Ross})$.
The outer boundary conditions 
to the internal structure in temperature and pressure
are taken at the Rosseland optical depth 20. In order
to obtain a smooth thermal gradient transition between the
atmosphere and the interior the thermal gradient is linearly interpolated 
with respect to $\rm \tau_{Ross}$ between its atmosphere value $\rm \nabla_a$ 
and its interior value $\rm \nabla_i$: so that 
for $\rm 20 < \tau_{Ross} < 1$ we consider
$\rm \nabla = x \nabla_i + (1-x) \nabla_a$ with $\rm x=\frac{\tau_{Ross}-1}{20-1}$.

{\bf Surface and 3-D convection simulations}

Owing to their calculations, we dub the thermal gradients 
coming from the 3-D simulations 
the \emph{hydrodynamic} and \emph{magnetohydrodynamic} gradients, 
$\rm \nabla_{HYD}$ and $\rm \nabla_{MHD}$ respectively.
These gradients are computed from the temperature and
gas pressure of the 3-D simulations. In 1-D calculations, the turbulent
pressure gradient is only accounted for in the equation of hydrostatic 
equilibrium. $\rm \nabla_{HYD}$ is obtained by horizontal and time
averages of 3-D surface simulations including no magnetic
field. Once we have ensured that 
this gradient is an average quantity
relevant for solar evolution from ZAMS to the actual Sun 
(see \S \ref{sec21}) we introduce it in the stellar structure 
equations. We do not expect that accounting
for the changes of surface conditions throughout
solar history  would greatly improve the results of this work. 
First these changes are moderate: the radius and the effective 
temperature increase by $\rm \approx 10\%$ and $\rm \approx 140 K$ respectively
from the solar ZAMS until today. 
More importantly the effects we investigate concern only 
the surface layers, they are not related to details of the inner structure 
resulting from the evolution of the Sun. Our results in \S \ref{sec31}
confirm that this is indeed the case. The average 3-D 
thermal gradient is introduced straightforwardly by considering
that the temperature optical depth relation is 
$\rm \frac{dT}{dr}=\nabla_{HYD}\frac{T}{P}\frac{dP}{dr}$ 
(with T the temperature, r the radius and, P the gas pressure) 
between the Rosseland optical depths $10^{-4}$ and $10^3$.
For $\rm \tau_{Ross}$ between $10^3$ and $10^{4}$
we base the thermal gradient on a linear interpolation
with optical depth between $\rm \nabla_{HYD}$ and $\rm \nabla_{CGM}$:
$\rm \nabla = x \nabla_{HYD} + (1-x) \nabla_{CGM}$ with 
$\rm x=\frac{10^4-\tau_{Ross}}{10^4-10^3}$ and $\rm \nabla_{CGM}$ the
gradient computed in the phenomenology of \cite{cgm96}.
We compute that at $\rm \tau_{Ross}=10^3$ ($\rm \approx 330\,km$ 
below the surface) the relative 
difference between $\rm \nabla_{HYD}$ and the adiabatic 
gradient $\rm \nabla_{ad}$ is about 15 percent (but much
larger above as can be seen from Figure~\ref{fig4}).
Thus most of the superadiabatic effects are expected above 
this depth. At the same time the transition between the 
dynamical description of the temperature gradient and the 
CGM description is shallow enough to enable us to adjust
the solar radius at solar age which is essential for the
purpose of our work (see \S \ref{sec23}). The model buildt that way
is called Solhyd. To investigate the 
surface magnetic effects on evolution we also build 
a model called Solmhd similar to Solhyd but where 
$\rm \nabla_{MHD}$ is used instead of $\rm \nabla_{HYD}$. 
Solhyd is calibrated following the
procedure described in the next section.

\subsection{Calibration and models surface properties}\label{sec23}

We consider that for the current Sun
$\rm R_{\odot}=6.9566\,10^{10} cm$, $\rm L_{\odot}=3.846\,10^{33} erg.s^{-1}$
and a metal to hydrogen mass fraction $\rm Z/X=1.81\,10^{-2}$. 
In order to obtain these 
quantities in the models after 4.6 Gyr of evolution, 
we adjust their initial $\rm Z/X$, helium fraction and convective
characteristic length scale. The chosen current radius 
obviously is a crucial parameter as of the 
calculation of eigenfrequencies.
There are different definitions possible of it.
What we consider here as the solar radius
is the radius of the layer at effective temperature.
Then $\rm R_{\odot}=6.9566\,10^{10} cm$ is the value recommended 
by \cite{haber08}. It lies $\rm 330 km$ below the commonly used
value $\rm R_{\odot}=6.9599\,10^{10} cm$ (\cite{allen76}) 
and nearly corresponds to the seismic
solar radius computed from f-modes whereas the \cite{allen76}
radius agrees with the position of the inflexion point to the
intensity profile at solar limb.
The calibration is achieved to better than $10^{-6}$ in radius, 
luminosity and $\rm Z/X$. It is important to achieve such
a high accuracy for the radius. The observed relative shifts 
of an eigenfrequency $\rm \nu$ between the solar maximum and 
minimum activity are the order of $\rm \Delta \nu / \nu =10^{-4}$.
The eigenfrequency of a mode of radial order n depends on the 
stellar radius following $\rm \nu_n \approx n (2\int_0^{R_{\odot}} \frac{dr}{c_s})^{-1}$, 
with $\rm c_s$ the sound speed.
Thus radii calibrated to much better 
than $\rm 10^{-4}$ are necessary to make sure that the 
differences calculated are not related to differences in the 
calibration radii. The solar model named Solmlt is based on the MLT.
We obtain $\rm \alpha_{MLT}=2.45$ for it. 
This rather large value is required to achieve the actual solar 
radius because of the low $\rm \alpha_{MLT}=0.5$ in the
atmosphere. Also considering $\rm \alpha_{MLT}=0.5$
in the atmosphere \cite{samadi06} found $\rm \alpha_{MLT}=2.51$.
For the model Solcgm, based on the CGM phenomenology, 
we have $\rm \alpha_{CGM}=0.785$.

\begin{figure}
\centering
\includegraphics[angle=90,width=8cm]{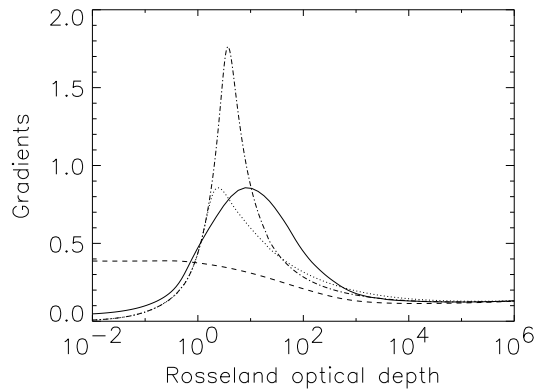}
\caption{Thermal and adiabatic gradients in the superadiabatic
layers of solar models as a function of the Rosseland
optical depth. Solid line: average thermal gradient from 3-D
simulations, dotted line: thermal gradient from the MLT phenomenology,
dot-dashed line: thermal gradient from the CGM phenomenology, 
dashed line: adiabatic gradient. \label{fig4}}  
\end{figure}

For a model relying on the 3-D surface simulations, 
it is also possible to adjust the final radius, luminosity 
and composition at the same time provided
the transition between the hydrodynamic thermal
profile and the phenomenological one is not performed
at too large an optical (or equivalently geometrical) 
depth. As mentioned at the end of the preceding section, the average
3-D temperature gradient is used down to $\rm \tau_{Ross}=10^3$.
Below $\rm \tau_{Ross}=10^4$ we rely on the CGM phenomenology. 
A linear interpolation is performed between the two 
descriptions at intermediate optical depths
(cf. \S \ref{sec22}). Doing so, almost all of the superadiabatic
convection region is treated
according to our 3-D prescriptions (see Figure~\ref{fig4}).
A smaller part of the superadiabatic region is
still handled with the CGM. It is significant
enough so that we keep a sufficient leverage on the deep 
convection zone specific entropy thanks 
to $\rm \alpha_{CGM}$ and therefore on the solar model
radius. This is how Solhyd is buildt.

\begin{figure}
\centering
\includegraphics[angle=90,width=8cm]{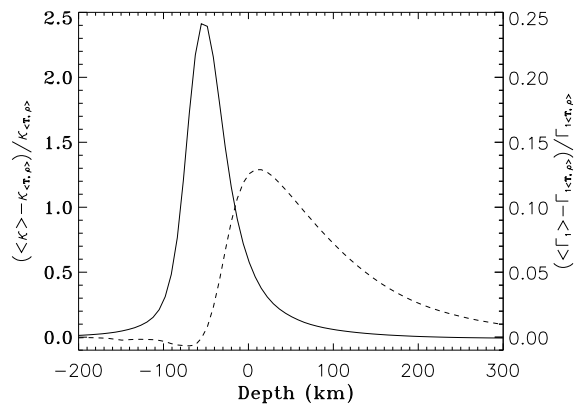}
\caption{Solid line : relative variation between averaged Rosseland opacity and 
Rosseland opacity of averaged density and temperature as a function
of depth. Dashed line : relative variation between averaged $\rm \Gamma_1$ 
adiabatic index and $\rm \Gamma_1$ adiabatic
index of averaged density and temperature as a function
of depth. Note the different scales for the Rosseland opacity and
the adiabatic exponent. \label{fig5}}  
\end{figure}

It is worth spending a few words on a model where the superadiabatic
convection completely relies on 3-D simulations. 
If we perform the transition between 
the 3-D thermal gradient and the phenomenological CGM one
between $\rm \tau_{Ross}=10^6$ and $\rm \tau_{Ross}=4\,10^6$ 
(near the lower boundary of the 3-D simulation box) the
final radius becomes nearly insensitive to $\rm \alpha_{CGM}$.
In such a {\it full 3-D convection} model changing $\rm \alpha_{CGM}$ 
from its solar calibration value 0.785 to 7.85 
decreases the radius by $\rm 0.3\%$ only.
As a comparison, in the models exclusively based on 
phenomenologies, the dependency on $\alpha$s is much larger : 
an increase of $\rm \alpha$ by 1 from the calibrated values reduces the 
radius by 6.3 and 8.5 percent in MLT and CGM models respectively.
Unsurprisingly the radius of the full 3-D convection model
departs from the solar radius at solar age.
It is $\rm \approx 1.3\%$ above the actual 
solar radius. Considering that
$\rm \alpha_{CGM}$ has no calibration effect on the model such a final
radius is fairly close to the solar radius. Yet 
the difference is enough to shift the eigenfrequencies by more
than 100 $\rm \mu Hz$ for modes of low-$\ell$ and radial 
order 15 to 30 thus masking the possible improvements brought
by the implementation of 3-D effects. 

Even though the thermal gradient and turbulent pressure 
accounted for in 1-D are estimated from 3-D simulations 
there are dimensional effects the 1-D cannot reproduce.
As illustrated on Figure~\ref{fig5} the average 
subphotospheric opacity is significantly
above the opacity corresponding to the average temperature
and density (see R99; \cite{nsa09}). 
This stems from the opacity's strong dependence
on temperature when $\rm H^{-}$ ions are its main contributor :
namely $\rm \kappa \propto T^{10}$. The horizontally averaged
opacity being larger in 3-D than in 1-D the subphotospheric 
temperature will be larger for the same total flux.
The dimensional effects are to a lesser extent seen on
EoS quantities such as the first adiabatic 
index $\rm \Gamma_1$ also shown on Figure~\ref{fig5}. 
To take into account such 3-D effects {\it directly} we
patched to 1-D solar structures the pressure, density, $\Gamma_1$ and
Brunt-V\"{a}iss\"{a}l\"{a} frequency average 
vertical profiles from 3-D simulations. 
The procedure is similar to that adopted by R99
and is repeated for purely hydrodynamic
and magnetohydrodynamic simulations in the models we
called Solpatchs.

\begin{table*}
  \caption[]{Main properties of the models.}
   \centering
     \label{Tabmod1}
 $$ 
     \begin{array}{lccccccccc}
        \hline
        \noalign{\smallskip}
        \rm  Name     &\rm \nabla-type\,along\,evolution& \rm Temperature      &  \rm Turbulent      &  \rm Magnetic  & \rm R_{T=5777}-R_{\odot} (km) & \rm R_{\tau=1}-R_{\odot} (km)  \\
                      &                               & \rm and\,density     &  \rm pressure       &  \rm field     &                           \\
        \noalign{\smallskip}                                                                 
        \hline                                                                               
        \noalign{\smallskip}                                                                 
        \rm Solmlt    &  \rm \nabla_{MLT}   & \rm No\, 3-D\,effects  &  \rm No          &    \rm No        &  -5.85\,10^{-1} &   -3.31 \,10^{1}   \\

        \rm Solcgm    &  \rm \nabla_{CGM}   & \rm No\, 3-D\,effects  &  \rm No          &    \rm No        &   2.65\,10^{-1} &   -3.31 \,10^{1}   \\
 
        \rm Solhyd    &  \rm \nabla_{HYD}   & \rm No\, 3-D\,effects  &  \rm Yes         &    \rm No        &   6.50\,10^{-1} &   -2.64 \,10^{1}   \\

        \rm Solmhd    &  \rm \nabla_{MHD}   & \rm No\, 3-D\,effects  &  \rm Yes         &    \rm Yes       &   7.21\,10^{3}  &    7.18 \,10^{3}   \\

        \rm Solpatch1 &  \rm \nabla_{HYD}   & \rm 3-D\,effects       &  \rm Yes         &    \rm No        &  -3.91\,10^{-1} &   -2.07 \,10^{1}   \\

        \rm Solpatch2 &  \rm \nabla_{HYD}   & \rm 3-D\,effects       &  \rm Yes         &    \rm Yes       &  -6.07\,10^{0}  &   -3.01 \,10^{1}   \\
 
        \rm Solpatch3 &  \rm \nabla_{MLT}   & \rm 3-D\,effects       &  \rm Yes         &    \rm No        &  -1.87\,10^{1}  &   -3.89  10^{1}    \\
 
        \rm Solpatch4 &  \rm \nabla_{MHD}   & \rm 3-D\,effects       &  \rm Yes         &    \rm Yes       &  -3.49\,10^{1}  &   -5.77 \,10^{1}   \\

        \noalign{\smallskip}
        \hline
     \end{array}
 $$ 
\end{table*} 

Table\ref{Tabmod1} sums up the properties of the models
we address in the remainder of this work. Col. 1 
gives the model name, Col. 2 gives the convective thermal
gradient used along solar evolution, Col. 3 indicates if 
the 3-D effects are accounted 
for directly (in the way mentioned just above), 
Col. 4 and 5 indicate respectively if 
turbulent pressure and magnetic fields are taken into account.
Col. 6 gives the differences between the radius of 
the model at $\rm T=5777 K$ and $\rm R_{\odot}$ (our reference
solar radius assumed to be here $\rm R_{\odot}=6.9566\,10^{10} cm$).
Col. 7 gives the differences between
the radius of the model at $\rm \tau_{Ross}=1$ and $\rm R_{\odot}$.
The models are calibrated in luminosity,
radius, and $\rm Z/X$ to better than $\rm 10^{-6}$. 
However as Solmhd, Solpatch2, Solpatch3 and Solpatch4 have
been buildt to explore the impact of the surface 
activity on the eigenfrequencies and the radius their
radii are not calibrated with a similar accuracy.
All the models have almost identical helium surface
mass fractions ranging from Y=0.2349 to 0.2352,
and also almost identical extent of the convection 
zone with radii of the base of the convection 
zone ranging from 0.7248 to 0.7249 $\rm R_{\odot}$. Let us finish
this section with some important remarks on the models :


i) The near surface stratifications 
computed in 3-D are more extended than those in 1-D with the MLT 
or the CGM phenomenology.
This is due to the larger subphotosperic temperatures
required in 3-D than in 1-D atmosphere models to produce the same
effective temperature (See R99 
for a discussion of this effect) and to an additional lift 
provided by the turbulent pressure support. 
Thus the Solpatch1 model is not buildt from Solhyd directly
but from a solar model
evolved with the same outer convection prescription
as Solhyd and whose photospheric radius is 115 km smaller
than our calibration value. This procedure allowed us to adjust
very accurately the Solpatch1 radius model to the 
solar radius.
In Solpatch1 the outer pressure, density, $\Gamma_1$ and
Brunt-V\"{a}iss\"{a}l\"{a} frequency vertical profiles are horizontal 
and time average of the 
purely hydrodynamic simulations down to our deepest layer of
the 3-D hydrodynamic simulation at $\rm T=2.225\,10^4$K, 
$\rm \rho=2.657\,10^{-5} g.cm^{-3}$ and $\rm \tau_{Ross}=1.88\,10^{7}$. 
Therefore the 3-D effects are directly accounted for instead of 
being introduced indirectly through $\rm \nabla_{HYD}$ as in Solhyd. 
We stress that this patch procedure only concerns the solar age model,
and not the solar evolution from ZAMS.
The Solpatch3 model is buildt exactly as Solpatch1 but from
a structure evolved with the MLT to a final radius
115 km smaller than Solmlt.

ii) The procedure to build Solpatch2 and Solpatch4 is 
similar to that for Solpatch1 except that the averaged surface
vertical profiles from the magnetohydrodynamic simulations
were used instead of the purely hydrodynamic profiles.
These averages are considered down to our deepest layer of
3-D magnetohydrodynamic simulation at $\rm T=2.225\,10^4$K,
$\rm \rho=2.647\,10^{-5} g.cm^{-3}$ and $\rm \tau_{Ross}=1.87\,10^{7}$.
Both Solpatch1 and Solpatch2 result from the same evolutionary 
sequence based on $\rm \nabla_{HYD}$. However for Solpatch1 the patch
at solar age is buildt from the hydrodynamic simulation whereas 
for Solpatch2 it is buildt from the magnetohydrodynamic one. 
Solpatch2 accounts for the 
magnetic effects around the actual solar surface but not for 
their feedback on the deeper structure during the solar evolution. 
On the contrary for Solpatch4, the structure is taken
from a solar evolution model evolved with $\rm \nabla_{MHD}$ and
the patch at solar age is made from the magnetohydrodynamic simulation.
Therefore Solpatch4 accounts for both the 
magnetic effects on the current solar structure and during the
solar evolution.

iii) The only difference between Solmhd and Solhyd are the 
thermal gradient and turbulent pressure. 
The purpose of Solmhd is to show the secular impact of 
magnetism on radius in the context of this work. 
Therefore Solmhd is not forced to solar
radius at solar age by performing a calibration.

iv) Regarding the (purely) hydrodynamic modelling 
our approach resembles the approach
of R99 and it is interesting to compare them.
For instance both studies investigate how the change from MLT
models to patched models affects the differences between
the predicted and the observed eigenfrequencies. These 
differences stem from the incorrect modelling of the 
outer average structure of the Sun, what is referred
to as the model effects. They also stem from the 
complex coupling of the modes with the radiative 
and convective fields, what is referred to as the 
modal effects. Contrary to R99 we don't investigate
both model and modal effects but only the model 
effects. Moreover our study is extended to models 
R99 did not address: the CGM model of convection (Solcgm)
and the models based on realistic thermal gradients (Solhyd, Solmhd).
Finally, R99 based most of their work on envelope models
whereas we calculate the oscillations from
models of the whole solar structure. For this series
of reasons, the present study is complementary 
to the study of R99.

\section{Results}\label{sec3}

We compute the eigenfrequencies of the previous
models using the adipls package (made publicly available by J. Christensen-Dalsgaard from
\verb=http://users-phys.au.dk/jcd/adipack.n/)=. Wave propagation is assumed adiabatic. 
We compare our theoretical frequencies to the Michelson
Doppler Imager (MDI) observations described by \cite{schou99}.
The data have been retrieved from \verb=http://quake.stanford.edu/~schou/anavw72z/=.
We consider two sets of data: Smax, the first set,
starts on August 27th 2001 and Smin, the second set, starts on December 12th 2008.
Each set lasts 72 days. 
Smax is representative of solar maximum activity and Smin of  
solar minimum activity. For every model there are $\approx$ 150 modes for which
we also have observational data. The orders go from 7 to 27,
the angular degrees from 0 to 10 and the frequencies from
from $\approx$1500 to $\approx$4300 $\mu$Hz.

\subsection{Dynamical and phenomenological models}\label{sec31}

\begin{figure}
\centering
\includegraphics[angle=90,width=8cm]{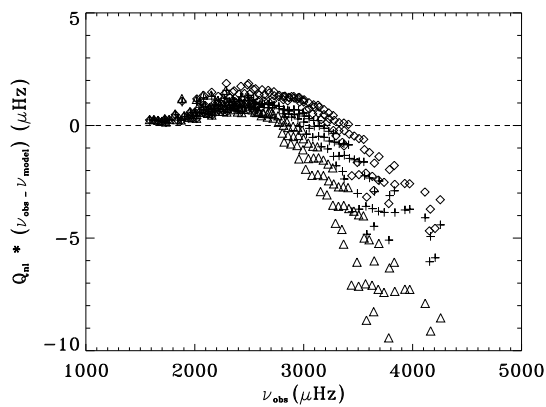}
\caption{Differences between models and observed eigenfrequencies
as a function of the observed eigenfrequencies. + : model 
Solmlt. $\rm \diamond$ : model Solcgm. $\rm \triangle$ : model Solhyd. The average
errobar on observed frequencies is $\rm \approx 0.06\, \mu Hz$ which is smaller 
than the symbol sizes.
 \label{fig6}}  
\end{figure}
\begin{figure}
\centering
\includegraphics[angle=90,width=8cm]{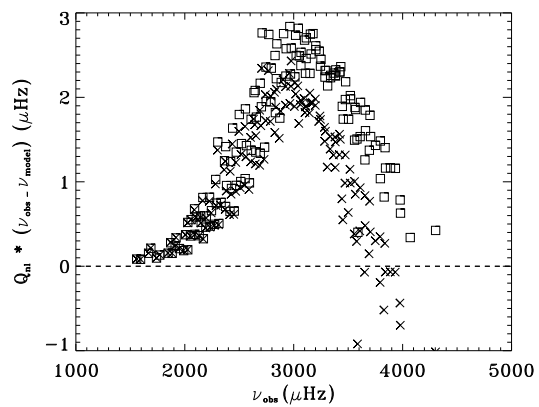}
\caption{Differences between models and observed eigenfrequencies
as a function of the observed eigenfrequencies. $\rm \square$ : model Solpatch1. 
$\rm \times$ : model Solpatch2. The Solpatch1 model is compared to
the Smin eigenfrequencies set representing the solar minimum whereas the
Solpatch2 model is compared to the Smax eigenfrequencies set representing the 
solar maximum. 
}  
\label{fig7}
\end{figure}

Figures \ref{fig6} and \ref{fig7} show the differences between
model and observed frequencies. As is customary the differences have
been scaled by $\rm Q_{nl}$ the ratio of the mode mass by the 
mode mass of the radial mode of same order. As the models we compare mainly 
differ in their surface layers, and we only
compare lower degree modes, the scaled frequency differences are expected
to be predominantly functions of frequency.

Figure \ref{fig6} shows the effects of changes in 
the treatment of thermal gradient and the turbulent 
pressure in the superadabatic region.
The modelled eigenfrequencies agree with the observations up
to $\rm \sim 3000\,\mu$Hz. Then they become 
larger than the observed ones with discrepancies increasing towards higher 
frequencies. At $\rm \sim 4000 \,\mu$Hz the 
MLT convection model and the CGM model exhibit frequencies about
5\,$\mu$Hz larger than the observed ones and
the frequencies of the model based on $\rm \nabla_{HYD}$ 
differ from the observed ones by about $\rm 8 \,\mu$Hz. 
These results suggest that the differences between observed and 
theoretical frequencies are not due to an incorrect calculation 
of the average thermal gradient. Solmlt and Solcgm 
exhibit quite different gradients (see Figure \ref{fig4})
but have similar eigenfrequencies. In spite of 
a gradient probably closer to the real 
one than Solmlt or Solcgm, the Solhyd model has eigenfrequencies 
that are not in better agreement with the observations.
This suggests that the ability of a model
to predict the correct eigenfrequencies is not only
related to its ability to reproduce the correct
average thermal gradient and the turbulent pressure. Figure \ref{fig6} and 
Figure 1 of R99 compare the differences between the 
eigenfrequencies of models 
based on the MLT and observed eigenfrequencies.
Both plots show that the predictions agree with 
observations at low frequency and become 
increasingly too large above a certain 
threshold. The thresholds differ. When focusing on the low $\ell$ p-modes
of Figure 1 of R99 (corresponding to the modes we compute)
we note that our MLT model is in agreement 
with the observations over a larger frequency range. 
Namely, at $\rm 3000\,\mu\,Hz$
and $\rm 4000\,\mu\,Hz$ R99 report discrepancies of 
$\rm \sim 5\,\mu\,Hz$ and $\rm \sim 12\,\mu\,Hz$ respectively 
whereas we report discrepancies of $\rm \sim 1\,\mu\,Hz$
and $\rm \sim 5\,\mu\,Hz$ respectively. We do not 
know the origin of this improvement.


Figure \ref{fig7} compares models Solpatch1 and Solpatch2 
showing the improvement brought by direct
patchs of the 3-D simulations average structure.
Solpatch1 and Solpatch2 are compared to the Smin 
and Smax set of data respectively.
These models frequencies predictions 
are less than $\rm 3\, \mu$Hz {\it below} 
the observations at $\rm \sim 3000\,\mu$Hz. This is their largest departure 
from the observed frequencies. At higher frequency the difference
decreases below $\rm 1 \, \mu$Hz. 
This results stresses the importance of accounting for the 3-D
nature of surface convection and not only for the correct thermal gradients 
and turbulent pressure when it comes to estimating the solar/stellar
absolute oscillation frequencies. 
If here again we want to compare our results with R99, 
Figure \ref{fig7} will obviously correspond to Figure 6 
of these authors. The models that are addressed in both cases  
are made of patchs of 3-D simulations averages. Interestingly, 
the differences computed with the observed eigenfrequencies have 
similar shapes. For frequencies below $\rm \sim 4000\,\mu$Hz
we compute eigenfrequencies smaller than the observed ones
with a maximum difference of $\rm \sim 3\,\mu$Hz.
Similarly, for frequencies below $\rm \sim 3500\,\mu$Hz, R99
compute eigenfrequencies smaller than the observed ones.
The maximum difference they compute is larger than in our 
work: $\rm \sim 4\,\mu$Hz. Above $\rm \sim 3500\,\mu$Hz
the eigenfrequencies of the R99 work become larger
than the observed ones. Using our numerical tools we were
unfortunately unable to determine oscillation frequencies towards
the larger frequencies, but it is clear from Figure \ref{fig7}
that a similar trend as in R99 is expected.

Let us now review three possible causes of the remaining
differences between the eigenfrequencies of the patched models 
Solpatch1 and Solpatch2 and the observations: 
the assumption made to buildt the solar evolution models,
the solar radius that is considered and, the uncertainties 
on the physics of the modes.

\begin{figure}
\centering
\includegraphics[angle=90,width=8cm]{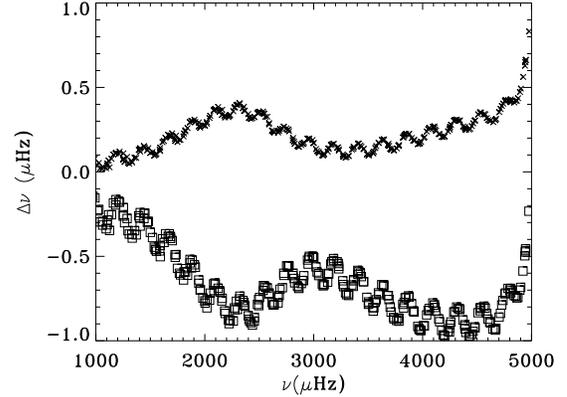}
\caption{Differences between models eigenfrequencies
as a function of their eigenfrequencies. $\rm \square$: 
model Solpatch1 minus model Solpatch3 frequencies. 
$\rm \times$ : model Solpatch2 minus model Solpatch4 frequencies.
 \label{fig8}}  
\end{figure}

First, how are our results sensitive to the hypotheses
on the surface convection adopted in the solar evolution models?
Figure \ref{fig8} shows that the type of
convection chosen to perform the past solar evolution has
a small impact on the seismic surface term of the present day Sun.
The models Solpatch1 and Solpatch3 correspond to patches of the 
same averaged atmosphere structure onto different solar 
interior structures. These structures correspond respectively to solar evolution
models using $\rm \nabla_{HYD}$ and $\rm \nabla_{MLT}$.
Yet Solpatch1 and Solpatch3 frequency differences are $\rm 1\,\mu$Hz at most.
As mentioned in \S \ref{sec22} this suggests that using a series of 
convection models to follow the varying surface conditions 
along the main sequence would probably 
make little difference for eigenfrequencies---at least in the solar case.
Figure \ref{fig8} also estimates how the eigenfrequencies change provided the surface 
magnetic activity is accounted for during the solar evolution.
In the Solpatch4 case, the evolution uses $\rm \nabla_{MHD}$.
The eigenfrequencies of this model are almost identical to
those of Solpatch2 model whose evolution uses $\rm \nabla_{HYD}$.

The models eigenfrequencies depend on the current global parameters
adopted and especially the radius.
How would a change of the calibration radius affect
the calculated eigenfrequencies?
This is an interesting question firstly in the 
perspective of asteroseismic investigations in other
solar-type stars for which the radius is known with
much less precision. Secondly because the value of the 
solar radius is not completely settled. For the solar effective temperature
layer, \cite{brown98} estimate 
$\rm R_{\odot}=6.9551\,10^{10} cm$ but it could be up to $\sim 300$ km
larger with $\rm R_{\odot}=6.9578\,10^{10} cm$ (\cite{antia98}).
Our calibration value stands inbetween at $\rm R_{\odot}=6.9566\,10^{10} cm$ but
we also computed the eigenfrequencies of a model
calibrated on a radius that is 300 km larger (and is therefore almost calibrated
on the \cite{allen76} radius). This model was buildt in the same manner as
Solpatch1 with the only difference of its radius. We found that 
its eigenfrequencies
are changed by $\rm 2\,\mu$Hz at most with respect to Solpatch1. 
Above $\rm 2500\,\mu$Hz the frequency shift induced by photospheric 
radius change is almost constant while it is negligible below 
$\rm 2000\, \mu$Hz.

Last but not least, the physics of the modes in the outer
layers is an obvious possible culprit of the observations/predictions 
differences remaining in Figure \ref{fig7}.
Our work however suggests that this contribution 
is not the major one. Indeed
R99 tried to approximate the modal effects whereas we
completely ignore them and nevertheless our calculations
show a better agreement to the observations. 
Thus we tend to think that a large part of the 
lingering difference between models and observations 
does not originate from the modal effects 
but from other physical processes to be included
in the model effects: the rotation, the large scale convective
and meridional flows, the role of the magnetic field that is present
even in the quiet Sun.

\subsection{Activity impact}\label{sec32}

In our magnetohydrodynamic run the induction equation
is solved throughout all the domain and simulation duration.
The plasma entering the simulation box from below carries 
a 1.2 kG horizontal magnetic field. This configuration is
inspired from \cite{baldner09} as
their helioseismic analysis suggests that 
the toroidal magnetic field at $\rm 0.996\,R_{\odot}$ (i.e. 
3000 km below the surface) is about 1.2 kG at maximum solar 
activity. The simulated magnetic field has an impact on both the
absolute radius of the Sun and its seismic properties.

\begin{figure}
\centering
\includegraphics[angle=90,width=8cm]{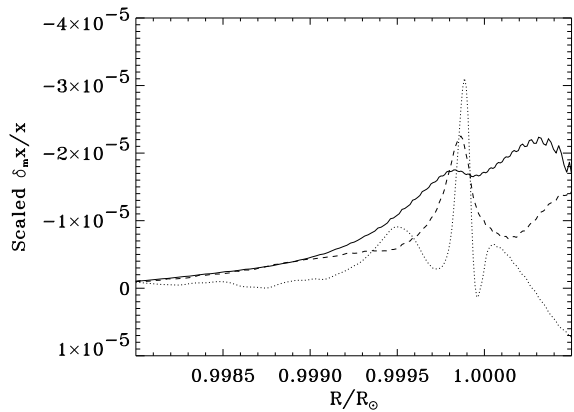}
\caption{Relative Lagrangian variations $\rm \delta_m x /x =(x_{MHD}-x_{HYD})/x_{HYD}$
as a function of radius between models Solpatch1 and Solpatch2. Solid line: radius variation, 
dashed line: sound speed variation times $\rm -2\,10^{-3}$, dotted line: turbulent pressure variation times 
$\rm  5\,10^{-4} \nabla P_{turb}^{HYD}/\nabla P_{gas}^{HYD}$.  
\label{fig9}}  
\end{figure}

\begin{figure}
\centering
\includegraphics[angle=90,width=8cm]{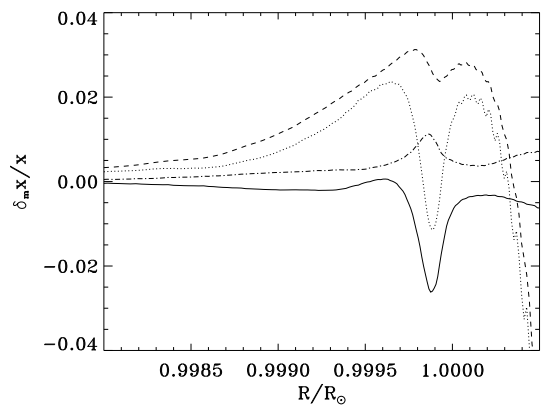}
\caption{Relative Lagrangian variation $\rm \delta_m x /x =(x_{MHD}-x_{HYD})/x_{HYD}$
as a function of radius between models Solpatch1 and Solpatch2. Solid line: $\rm \upsilon$, 
dotted line: density, dashed line: pressure, dot-dashed line: sound speed.
\label{fig10}}  
\end{figure}

Let us first examine the activity impact on the radius.
In the following the relative MHD/HYD variation considered is  
$\rm \delta x /x =(x_{MHD}-x_{HYD})/x_{HYD}$ for any quantity x. 
Figure \ref{fig9} illustrates the Lagrangian 
relative radius variation $\rm \delta_m r /r$ (i.e.
the relative radius variation throughout layers corresponding to the same 
mass coordinate) between the purely hydrodynamic 3-D (Solpatch1) 
and the magnetohydrodynamic 3-D simulations (Solpatch2).
The radius variation is 
most significant in the $\rm \sim 900 km$ region extending above 
the solar photosphere. Here we are, however,
only concerned in the solar interior.
For the solar interior the contraction is most significant 
with $\rm \delta_m r /r \sim -2\,10^{-5}$ at the surface and 
within the first 200 km below it. The shrinking clearly corresponds 
to a sharp drop in the turbulent pressure support. 
The turbulent pressure gradient also changes  
deeper than 200 km, yet in those underlying regions the ratio of 
turbulent pressure gradient to gas pressure gradient becomes 
negligible and so are the effects of its change (see also
Eqs. 12 and 15 of \cite{trampedach13}). 
For this reason we have drawn in Figure \ref{fig9} the relative
variation of the vertical turbulent pressure gradient when 
the magnetic field is accounted for
($\rm (\nabla P_{turb}^{MHD}-\nabla P_{turb}^{HYD})/ \nabla P_{turb}^{HYD}$)  
multiplied by the ratio of turbulent pressure 
gradient to gas pressure gradient 
($\rm \nabla P_{turb}^{HYD}/\nabla P_{gas}^{HYD}$).
The Figure also shows the impact of magnetism 
on sound speed. Right beneath the surface the sound 
speed $\rm c_s$ increases the order 
of $\rm 1\%$. In order to keep the same ordinate axis 
scale for all the quantities shown on Figure \ref{fig9}, 
the turbulent pressure gradient and the sound speed
variations have been scaled by $\rm 5\,10^{-4}$ and $\rm -2\,10^{-3}$ 
respectively. The correlation between the variation of turbulent pressure support
and radius and the anti-correlation between the variation of turbulent pressure 
support and sound speed are striking.
Figure \ref{fig10} shows the relative variations of pressure,
density, sound speed $\rm c_s$, and $\rm \upsilon=\frac{\Gamma_1}{c_s}$,
$\rm \Gamma_1$ being the first adiabatic index.
Deeper than 1000 km below $\rm R_{\odot}$ the changes in any of
these quantities become rapidly negligible. 
Between 1000 and 200 km below $\rm R_{\odot}$ 
pressure and density increase altogether which results
in a nearly constant sound speed and $\rm \upsilon$. 
Finally from 200 km below $\rm R_{\odot}$ 
up to the surface $\rm c_s$ increases 
because of pressure increase and constant
density and in spite of a decreased first adiabatic 
index $\rm \Gamma_1$. $\rm \upsilon$ 
exhibits a sharp drop.

Before moving to the seismic effects
let us comment on the impact of activity on the observed radius. 
A stellar radius may be defined in many ways.
We have considered the solar calibration radius to be the radius of the
effective temperature layer but we report on the changes in both 
this radius and the radius of the $\rm \tau_{Ross}=1$ layer
\footnote{In the 3-D models we compute $\rm \tau_{Ross}$ as the 
average of optical depths over the $240^2$ cells
of similar geometrical depth} in Table \ref{Tabmod1}. 
From the model Solpatch1 to Solpatch2 we 
compute that the shells at effective temperature and $\rm \tau_{Ross}=1$
go down by respectively 5.7 km and 9.4 km. As Solpatch1 and Solpatch2 are
obtained from the same evolutionary sequence, we therefore estimate that
the immediate effect of the magnetic activity on the structure is to decrease
the radius by 8 milliarcseconds (hereafter mas) 
at the effective temperature layer and 14 mas at $\rm \tau_{Ross}=1$.
Note that the effective temperature layer lies at $\rm \tau_{Ross} \approx 0.55$ 
in the photosphere and that consequently we predict that the radius variation 
increases with depth in the visible part of the atmosphere.
The predicted radius changes will be difficult to detect
because they remain smaller
than the constant fluctuations of depth induced 
by convective overshooting or p-modes oscillations: the
order of 30 km for the effective temperature layer.
If the immediate effect of surface magnetism is to decrease the
radius we compute that its secular effect is on the contrary to 
increase it in agreement with other works 
(\cite{lydon95}; \cite{macdonald10}). 
There is no contradiction in this. Solmhd has 
a much larger radius than Solhyd because of a slightly larger
specific entropy of its deep convection zone.
The difference stems from $\rm \nabla_{HYD}$ and 
$\rm \nabla_{MHD}$: when we perform the downward
integration of these thermal gradients starting from the 
same surface conditions ($\rm T=5777$ K and $\rm \rho=2.5\,g.cm^{-3}$)
we end up on an adiabat that is $\rm \approx 1.96\,10^7 erg.K^{-1}.g^{-1}$ 
higher in specific entropy in the magnetohydrodynamic case
than in the purely hydrodynamic case. Magnetic fields
impede convective motions
(see \S \ref{sec21}) and eventually convection becomes less efficient.
Actually Solhyd and Solmhd deep convection zone specific entropies
differ only by $\rm \Delta s =7.90\,10^6 erg.K^{-1}.g^{-1}$ because as mentioned
in \S \ref{sec22} their thermal gradients exclusively rely on
$\rm \nabla_{HYD}$ and $\rm \nabla_{MHD}$ only down $\rm \tau_{Ross}=10^3$.
The time required for the solar luminosity to change
the specific entropy from the Solhyd value to the Solmhd value 
is $\rm \Delta s \int_{CZ}Tdm / L_{\odot} \approx 3500$ yr
where the integration is made over the convection zone with 
T the temperature, and dm the mass increment. Therefore
the duration of a cycle is much too short for the
change in the outer gradient to affect the thermal 
structure deeper down and eventually increase the radius.

\begin{figure}
\centering
\includegraphics[angle=90,width=8cm]{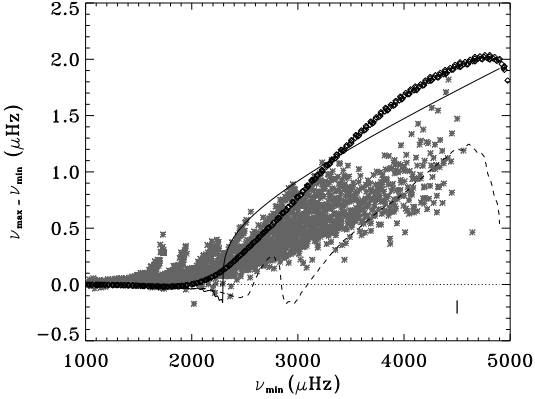}
\caption{$\rm \diamond$: frequency shift between models Solpatch1 and Solpatch2
as a function of frequency. Solid line: frequency shift computed
from Equation \ref{eq1} and cut-off frequency from Equation \ref{eq3}, 
Dashed line: frequency shift computed from Equation \ref{eq1} and 
cut-off frequency from Equation \ref{eq2}. Dotted line: 
frequency shift computed from the Equation \ref{eq1} 
$\rm \frac{\delta_m r}{r}$ term. Grey crosses: observed frequency
shifts for p-modes. The average errorbar on frequency shifts 
$\rm 0.06\,\mu Hz$ is shown at 
the lower right part of the plot.
\label{fig11}}  
\end{figure}

\begin{figure}
\centering
\includegraphics[angle=90,width=8cm]{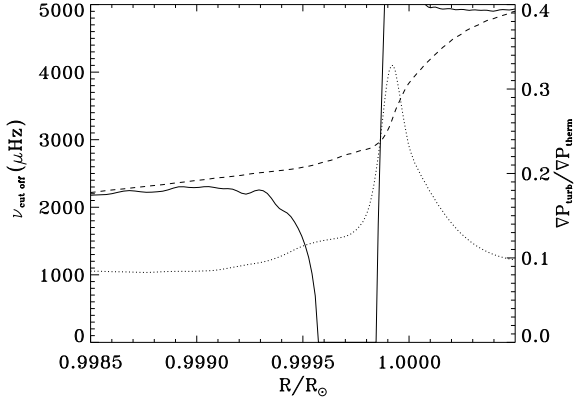}
\caption{Left hand ordinate axis: cut-off frequency from equation \ref{eq1} vs radius (solid line).
cut-off frequency estimate from equation \ref{eq2} vs radius (dashed line).
Right hand ordinate axis: turbulent pressure gradient to gas pressure gradient ratio vs radius (dotted line). 
All the physical quantities of this Figure are related to the Solpatch1 model.
\label{fig12}}  
\end{figure}

The frequencies of solar p-modes are observed to change with phase of
the solar activity cycle.
From minimum to maximum solar activity \cite{chaplin07}
and \cite{rabello08} report frequency increases
of 0.3 to 0.4 $\mu$ Hz around 3000 $\mu$Hz.
Figure \ref{fig11} shows the differences between 1928 modes
of the Smin and Smax datasets. The observed frequencies 
ranges from $\rm 1025\, \mu Hz$ to $\rm 4644\, \mu Hz$, the radial order 
from 0 to 27, and the degree from 0 to 298. 
In this data sample the activity related shifts increase with frequency
to reach $\rm 1\mu$Hz at 4000 $\mu$Hz.   
We drawn in overplot the eigenfrequency differences
between models Solpatch1 ($\rm \nu_{min}$) and Solpatch2 ($\rm \nu_{max}$). 
Note that the modal magnetic effects are not included in our
theoretical p-modes calculation, the changes in
eigenfrequencies only results from the changes in the
near surface structure.
The models predict no frequency variations below $\rm \sim 2200\, \mu Hz$
and actually they can predict no variation 
below $\rm \sim 1800\, \mu Hz$ as this is the cut-off frequency
at the  base of our hydrodynamic/magnetohydrodynamic simulations.
Small eigenfrequency changes the order of $\rm 0.3 \mu$Hz are seen 
in the data at low frequency. Magnetic activity effects extending deep in the 
convection zone could explain them (\cite{mullan07}). 
Above $\rm \sim 2200\, \mu Hz$, the activity related frequency shift
increases nearly linearly up
to $\rm \sim 4000\, \mu Hz$ : the slope is $\rm 10^{-3}\, \mu Hz / \mu Hz$.
The amplitude of the observed frequency shift is about two thirds of our
predictions. At $\rm 3000\, \mu Hz$ 
the average shift from the data we use is $\rm 0.51\, \mu Hz$ in the
$\rm 3000 \pm 50\, \mu Hz$ bin while our calculation gives $\rm 0.75\, \mu Hz$. 
Given the spread in the observed frequency shifts the models 
predictions are actually compatible with the upper envelope of them.
At frequencies higher than $\rm \sim 4000\, \mu Hz$
\footnote{We are here able to evaluate the impact 
of activity for frequencies higher than $\rm \sim 4000\, \mu Hz$ whereas 
in \S \ref{sec31} we could not perform the direct comparison to the 
observed frequencies above this value. The reason is that for the 
low order modes $\rm \ell \leq 10$ whose frequency we can compute 
we have no corresponding MDI data above $\rm \sim 4300\, \mu Hz$.} the difference 
reaches a maximum of $\rm 2\, \mu Hz$ at $\rm \sim 4800\, \mu Hz$ and then
shows a slight decline.

\cite{Jcd97} have related 
the changes in their surface models properties to the variations of 
their eigenfrequencies through
\begin{equation} 
\begin{multlined}
\frac{\delta \nu}{\nu} \approx \Big( \int_0^{R_c} (1-\frac{\nu_c^2}{\nu^2})^{1/2} [\frac{\delta_mr^2P}{r^2P}-2\frac{\nu_c^2}{\nu^2}(1-\frac{\nu_c^2}{\nu^2})^{-1}\frac{\delta_mr}{r}]\frac{dr}{c_s}   
\\
{+ \int_0^{R_c} (1-\frac{\nu_c^2}{\nu^2})^{-1/2}\frac{\delta_m\upsilon}{\upsilon}\frac{dr}{c_s} \Big) / \int_0^{R_c} (1-\frac{\nu_{c}^2}{\nu^2})^{-1/2}\frac{dr}{c_s}}
\end{multlined}
\label{eq1}
\end{equation}
with $\rm \delta \nu / \nu$ the relative frequency variation, $\nu_c$ 
the cut-off frequency, P the pressure, r the radius, 
$\rm c_s$ the sound speed, and $\rm \upsilon=\Gamma_1/c_s$.
$\rm \delta_m$ stands for the Lagrangian variations of the physical 
quantities and $\rm R_c$ for the cut-off radius.
Using this equation we will now outline some interesting features
in our models. The dotted line in Figure \ref{fig11}
shows that we expect no visible eigenfrequency changes
if the only term taken into account in Equation \ref{eq1}
is $\rm \frac{\delta_mr}{r}$. This could be anticipated as
with $\rm \frac{\delta_m r}{r} = 1.7\,10^{-5}$ at most in the interior (Figure \ref{fig9}), 
the radius variations are tiny.
We find that the largest variation of the cut-off radius 
occurs for $\rm \nu_c = 5000 \, \mu Hz$
and is $\rm \frac{\delta R_{c}}{R_{c}} = 5\,10^{-6}$.
These results suggests that {\it the frequency variations are not 
related to variations of the
radius of the acoustic cavity but are related to changes inside
the acoustic cavity}: changes in pressure and $\rm \upsilon$.

The dashed line in Figure \ref{fig11} shows the eigenfrequencies
variations calculated from Equation \ref{eq1} provided the
cut-off frequency $\nu_c$  is computed using the pressure
scale height $\rm H_p$ as is frequently done:
\begin{equation} 
\rm 2 \pi \nu_{c} \approx \frac{c_s}{2H_{p}}
\label{eq2}
\end{equation}
The shape of the line is directly related to the shapes
of $\rm \frac{\delta_m\upsilon}{\upsilon}$ and $\rm \frac{\delta_mP}{P}$ seen 
on Figure \ref{fig10}. In particular the change of
sign in the predicted eigenfrequency variation at 
about $\rm 2800\,\mu$Hz is related to the marked drop
in $\rm \frac{\delta_m\upsilon}{\upsilon}$ that can be
seen at $\rm 0.9998\,R_{\odot}$ in Figure \ref{fig10}. 
Figure \ref{fig12} shows that the cut-off frequency
computed from Equation \ref{eq2} nearly is $\rm 2800\,\mu$Hz at $\rm 0.9998\,R_{\odot}$.

Using the pressure scale height in cut-off frequency 
calculation actually is an approximation. The
correct formula for cut-off frequency is based on
$\rm H_{\rho}$, the density scale height (\cite{Deubner84}):
\begin{equation} 
\rm 4 \pi^2 \nu_{c}^2=\frac{c_s^2}{4H_{\rho}^2}(1-2\frac{dH_{\rho}}{dr})
\label{eq3}
\end{equation}
The solid line in Figure \ref{fig11} shows the eigenfrequencies
variations calculated from Equation \ref{eq1} using this
cut-off frequency.
In this case there is no sign change in the eigenfrequency 
variation as in the preceding case. There are two reasons for this.
First, as shown in 
Figure \ref{fig12} the cut-off frequency computed from
Equation \ref{eq3} becomes zero on the $\rm[ 0.9996,0.9998]\,R_{\odot}$
interval. Consequently, the eigenfrequency variations
are less sensitive to the $\rm \frac{\delta_m\upsilon}{\upsilon}$ and 
more sensitive to $\rm \frac{\delta_mP}{P}$ changes than 
when the Equation \ref{eq2} is used for the cut-off frequency.
Second, the acoustic cavity is smaller than $\rm 0.9999\,R_{\odot}$ 
at any frequency whereas it clearly extends above $\rm 1.0005\, R_{\odot}$
when the Equation \ref{eq2} is used for cut-off frequency.
Figure \ref{fig11} shows that the calculation based on the
correct cut-off frequency is in better
agreement with the observations (or the direct calculation
of variations) than the calculation based on the approximated cut-off frequency. 
In the context of activity related shifts one 
should not use the pressure scale height to calculate an approximate cut-off frequency
because it strongly differs from the exact cut-off frequency (Figure \ref{fig12}).
There is an interesting consequence to this as, provided the Equation \ref{eq3} 
is used for cut-off frequency 
calculation, the eigenfrequencies variations become mostly 
sensitive to pressure changes.









\section{Conclusion}\label{sec4}

We have addressed the effects of surface convection and magnetism modelling on the
solar models p-modes frequencies and radii. We used a 1-D code to follow the 
evolution of the whole Sun from ZAMS until now and a 3-D code
to simulate its current magnetoconvection over 4 Mm around the surface. 
Different prescriptions for the surface convection were adapted to the stellar 
evolution code: the MLT, the CGM phenomenology, and averaged stratifications 
from our 3-D simulations. In our magnetic simulation runs
a 1.2 kG horizontal field is carried by the incoming fluid at the
lower boundary of the domain. This configuration mimics the 1.2 kG toroidal field
suggested by the seismic analysis of \cite{baldner09} at $\rm 0.996 R_{\odot}$
and for maximum activity. We compared the eigenfrequencies of the models to recent observations for 
about 150 eigenfrequencies. We also discussed the changes in modelled radius 
with activity. Our results are as follows:

{\bf Convection hydrodynamics and seismology}

For any of the adopted surface convection the models predict 
the solar p-modes eigenfrequencies within $\rm 2 \,\mu$Hz 
up to $\rm \approx 3000 \,\mu$Hz. The situation changes above that value:
both the MLT and the CGM phenomenologies show comparable 
trends to overestimate the eigenfrequencies. 
The CGM phenomenology provides a slightly better 
agreement with the observations because its superadiabatic 
region is very narrow and the superadiabaticity very intense.
Because of that, the turbulent pressure of the CGM model
expands the acoustic cavity which leads to 
frequencies in slightly better agreement with
the observations than in the MLT case. 
However the CGM phenomenology thermal gradient 
is much larger than what is suggested by the 3-D calculations while
the MLT superadabatic gradient is comparable to them.
When the subsurface thermal gradient is adapted from the averaged 3-D calculations 
and the turbulent pressure is accounted for, the predicted frequencies are 
higher than those of the models relying on the MLT or the CGM. 
The situation only improves if the surface layers pressure, density and first 
adiabatic index are directly taken from the averaged 3-D computations.
This is in good agreement with the results of \cite{rosenthal99}
who performed similar patchs. But, in addition to this previous work, it also 
outlines that accounting for the correct outer structure in 
the context of 1-D solar evolution does not produce better 
results than the usual phenomenologies.
In the so-called patched models the computed eigenfrequencies remain 
very close to and slightly {\it below} the observed ones in the frequency 
domain above 3000 $\rm \mu Hz$. 
The improvement brought by these patched models weakly depends on
the outer convection model (phenomenological or based on simulations) used to 
perform the preceding solar evolution to obtain the deeper solar structure. 
The hydrodynamic models that are 
patched do not account for magnetic activity. Accounting for it 
even at the moderate level of solar minimum would increase the eigenfrequencies 
and improve further the agreement with observations.
For the patched models, the mean pressure, mean first adiabatic index, mean density and,
mean energy are not related by the equation of state. 
Because of their non-linear
relationships the average of the local values is not equal to the equation of
state values of the means.
For the same reason, the average opacity 
is not the opacity computed from the average temperature and density.

{\bf Activity and radius}

For a given solar interior structure, when the surface layers are patched 
from the magnetohydrodynamic simulation instead of the hydrodynamic one
the effect is a $\rm 1.7\,10^{-5}$ Lagrangian radius relative decrease at
the surface and directly below. 
This corresponds to a $\rm \approx 12$ km shrinking of the mass layers
at surface but only translates into a $\rm \approx 6$ km shrinking of the 
effective temperature layer (or $\rm \approx 10$ km shrinking 
of the optical depth layer at $\rm \tau_{Ross}=1$). The difference 
stems from the Lagrangian relative variations of temperature and density near the
surface (that are considerably larger than $\rm 1.7\,10^{-5}$, see \cite{Jcd97}). They 
induce an opacity increase at a given mass coordinate 
which attenuates the Lagrangian radius decrease.
The radius of a layer at a given optical depth is related to 
both its geometrical depth and to the opacity variations of the overlying regions.
One should keep in mind that these results only 
concern the effects of activity modelled in the $\rm 4\,10^{-3} R_{\odot}=3 Mm$ 
region immediately below the surface. f-modes have indeed shown that radius 
variations occur in deeper regions and down to $\rm 0.97 R_{\odot}$ 
(\cite{lefebvre05}). Obviously these variations could 
also affect the visible radius. However the surface activity 
affects the atmosphere relative stratification and this is perhaps less sensitive
to the physics of the deeper regions. 
From the optical depth point of view our simulations suggest that the
radius contraction with activity should increase with depth on the 
$\rm 10^{-1} \leq \tau_{Ross} \leq 1$ range. For instance, we predict that the distance
between the layer at $\rm \tau_{Ross} \approx 0.55$ (T=5777 K) and the one at $\rm \tau_{Ross}=1$  
increases from 20 km with no activity to 24 km with activity: that is a 6 mas difference.

{\bf Activity and seismology}

We computed the changes in eigenfrequencies when the magnetic activity is
accounted for and compared them to the differences observed
between activity maximum and minimum. 
Our calculations reproduce the order of magnitude and frequency dependence 
of frequency changes along the Hale activity cycle. No variation is predicted 
below $\rm 2200\, \mu Hz$. From this point on, the shift increases almost linearly  
to reach $\rm 1.7\, \mu Hz$ at 4000 $\rm \mu Hz$. 
The frequency increase is not related to a change in the
size of the acoustic cavity but mostly to a thermal 
pressure increase in a $\rm 10^{-3}\,R_{\odot}$ layer 
starting right below the solar surface at $\rm R_{\odot}=6.5966\,10^5\,km$.
The observed frequency shifts are spread and our calculations
correspond nearly to the upper envelope of the observations. There
might be several reasons to the slight overestimation of frequency shifts. 
The simulated incoming magnetic field may be too strong. 
Secondly, we did not explore the impact of the poloidal component
of the magnetic field which would be introduced as a vertical component
of the incoming field in our geometry. In other words our
incoming field configuration is perhaps too simple.
Thirdly, the models we use for frequency variations are 1-D, 
we do as if the toroidal field
at $\rm 0.996 R_{\odot}$ had the same intensity whatever the latitude.
Fourthly, all the rotation effects such as the meridional flow are ignored 
and the simulation box is too small for the mesogranular or supergranular 
scales to be accounted for.
Finally, by comparing the purely hydrodynamic model to the magnetohydrodynamic 
one we completely ignore the magnetic field effects at the cycle 
minimum which leads to an overestimated minimum to maximum contrast.
It is clear that a lot more investigations need to be done.

3-D simulations of solar surface convection have been known for
some time to diminish the seismic surface term. This work
shows that above $\rm 2200\, \mu Hz$ the magnetic activity has a potential impact
of the order of a $\rm 1\, \mu Hz$ on the eigenfrequencies.
This is the order of the observed shifts along solar cycle.
This is also the order of the differences between the observed
frequencies and those computed from 3-D simulations without 
any magnetic field. Thus further efforts in 
explaining the solar surface term should now take into account
the effects magnetic activity even in moderately active 
stars such as our Sun.

\section*{Acknowledgements}
We thank the anonymous referee for significantly contributing to the improvement
of this work.


%

\label{lastpage}


\begin{thebibliography}{}

\bibitem[\protect\citeauthoryear{Allen}      {1976}]{allen76}           Allen, C. W., 1976, Astrophysical Quantities (3rd ed.; London: Athlone)
                                           
\bibitem[\protect\citeauthoryear{Angulo et al.}     {1999}]{angulo99}          Angulo, C., et al., Nucl. Phys. A656 (1999)3-187
                                           
\bibitem[\protect\citeauthoryear{Antia}      {1998}]{antia98}           Antia, H.M., 1998, A\&A 330, 336
                                           
\bibitem[\protect\citeauthoryear{Asplund et al.}       {2005}]{aspl05}            Asplund, M., Grevesse, N., Sauval, J., 2005, ASP Conference Series, Vol XXX.
                                           
\bibitem[\protect\citeauthoryear{Asplund et al.}    {2009}]{asplund09}         Asplund, M. Grevesse, N. Sauval, A. J, Scott, P., 2009, A\&RAA, 47, 981
                                           
\bibitem[\protect\citeauthoryear{Baldner et al.}    {2009}]{baldner09}         Baldner, Charles S., Antia, H. M., Basu, Sarbani, Larson, Timothy P., 2009, ApJ, 705, 1704
                                           
\bibitem[\protect\citeauthoryear{Basu et al.}       {1999}]{basu99}            Basu, S., Antia, H. M., Tripathy, S. C., 1999, ApJ, 512, 458
                                           
\bibitem[\protect\citeauthoryear{Beeck et al.}      {2012}]{beeck12}           Beeck, B., Collet, R., Steffen, M., Asplund, M., Cameron, R. H., Freytag, B., Hayek, W., Ludwig, H.-G., Schüssler, M., 2012, A\&A, 539, 121
                                           
\bibitem[\protect\citeauthoryear{Bohm-Vitense}       {1958}]{bohm58}            B\"{o}hm-Vitense, E., 1958, Zs. f. Ap., 46, 108
                                           
\bibitem[\protect\citeauthoryear{Brown \& Christensen-Dalsgaard}      {1998}]{brown98}           Brown, T. M., Christensen-Dalsgaard, J., 1998, ApJ, 500, 195
                                           
\bibitem[\protect\citeauthoryear{Canuto et al.}        {1996}]{cgm96}             Canuto, V. M., Goldman, I., Mazzitelli, I., 1996, ApJ, 473, 550 
                                     
\bibitem[\protect\citeauthoryear{Castelli}        {2005}]{castelli05}   Castelli, F., Mem. S.A.It. Suppl., 2005, 8, 25 
      
\bibitem[\protect\citeauthoryear{Chaplin et al.}    {2007}]{chaplin07}         Chaplin, W. J., Elsworth, Y., Miller, B. A., Verner, G. A., New, R., 2007, ApJ, 659, 1749
                          
\bibitem[\protect\citeauthoryear{Christensen-Dalsgaard et al.}        {1996}]{Jcd96}             Christensen-Dalsgaard, J., Nuclear Physics B, 48, 325
                 
\bibitem[\protect\citeauthoryear{Christensen-Dalsgaard \& Thompson}        {1997}]{Jcd97}             Christensen-Dalsgaard, J., Thompson, M.J., 1997, MNRAS 284, 527
                                           
\bibitem[\protect\citeauthoryear{Collet et al.}     {2011}]{collet11}          Collet, R., Hayek, W., Asplund, M., Nordlund, A., Trampedach, R., Gudiksen, B., 2011, A\&A, 528, 32
                                           
\bibitem[\protect\citeauthoryear{Corbard \& Thompson}    {2002}]{corbard02}         Corbard, T., Thompson, M. J., 2002, Sol Phys 205, 211
                                           
\bibitem[\protect\citeauthoryear{Deubner \& Gough}    {1984}]{Deubner84}         Deubner, F.-L., Gough, D., 1984, ARA\&A 22, 593
                                           
\bibitem[\protect\citeauthoryear{Feautrier}       {1964}]{feau64}            Feautrier, P., 1964, Comptes Rendus Acad. Sci. Paris, 258, 3189
                                           
\bibitem[\protect\citeauthoryear{Ferguson et al.}   {2005}]{ferguson05}        Ferguson, J. W., Alexander, D. R., Allard, F., Barman, T., Bodnarik, J. G., Hauschildt, P. H., Heffner-Wong, A., Tamanai, A., 2005, ApJ, 623, 585
                                           
\bibitem[\protect\citeauthoryear{Gustafsson et al.}      {1975}]{gusta75}           Gustafsson, B., Bell, R. A., Eriksson, K., Nordlund, A., 1975, A\&A, 42, 407
                                           
\bibitem[\protect\citeauthoryear{Haberreiter et al.}      {2008}]{haber08}           Haberreiter, M., Schmutz, W., Kosovichev, A. G., 2008, ApJ, 675, 53
                                           
\bibitem[\protect\citeauthoryear{Lefebvre et al.}   {2005}]{lefebvre05}        Lefebvre, S., Kosovichev, A., G., ApJ, 633, 149
                                           
\bibitem[\protect\citeauthoryear{Ludwig et al.}     {2006}]{ludwig06}          Ludwig, H.G., Allard, F., Hauschildt, P. H., 2006, A\&A, 459, 599
                                           
\bibitem[\protect\citeauthoryear{Ludwig et al.}     {2002}]{ludwig02}          Ludwig, H.G., Allard, F., Hauschildt, P. H., 2002, A\&A, 395, 99
                                           
\bibitem[\protect\citeauthoryear{Lydon \& Sofia}      {1995}]{lydon95}           Lydon, T., Sofia, S., 1995, ApJS, 101, 357
                                           
\bibitem[\protect\citeauthoryear{Macdonald \& Mullan}  {2010}]{macdonald10}       Macdonald, J., Mullan, D. J., 2010, ApJ, 723, 1599
   
\bibitem[\protect\citeauthoryear{Profitt \& Michaud}  {1993}]{proffitt93}      Proffitt, C. R., Michaud, G., 1993, ASP Conference Series, Vol. 40, 246
                                        
\bibitem[\protect\citeauthoryear{Montalban et al.}     {2004}]{montal04}          Montalban, J., D'Antona, F., Kupka, F., Heiter, U., 2004, A\&A, 416, 1081
                                           
\bibitem[\protect\citeauthoryear{Morel}      {1997}]{morel97}           Morel, P., 1997, A\&AS, 124, 597
                                           
\bibitem[\protect\citeauthoryear{Morel \& Lebreton}{2008}]{morelebreton08}     Morel, P., Lebreton, Y., 2008, Ap\&SS, 316, 61
                                           
\bibitem[\protect\citeauthoryear{Mullan et al.}     {2008}]{mullan07}          Mullan, D. J., MacDonald, J., Townsend, R. H. D., 2007, ApJ, 670, 1420
                                           
\bibitem[\protect\citeauthoryear{Nordlund et al.}        {2009}]{nsa09}             Nordlund, Ake, Stein, Robert F., Asplund, M., 2009, LRSP, 6, 2
                                           
\bibitem[\protect\citeauthoryear{Nordlund}       {1982}]{nord82}            Nordlund, A., 1982, A\&A, 107, 1

\bibitem[\protect\citeauthoryear{Piau et al.}       {2005}]{piau05} Piau, L., Ballot, J., \& , Turck-Chi\`eze, S., 2005, A\&A, 430, 571
                                           
\bibitem[\protect\citeauthoryear{Piau et al.}       {2011}]{piau11}            Piau, L. Kervella, P., Dib, S., Hauschildt, P., 2011, A\&A, 526, 100
                                           
\bibitem[\protect\citeauthoryear{Rabello-Soares}    {2008}]{rabello08}         Rabello-Soares, M. C., Korzennik, S. G., Schou, J., 2008, Sol. Phys., 251, 197
                                           
\bibitem[\protect\citeauthoryear{Robinson et al.}   {2003}]{robinson03}        Robinson, F. J., Demarque, P., Li, L. H.,Sofia, S., Kim, Y.-C., Chan, K. L., Guenther, D. B., 2003, MNRAS, 340, 923
                                           
\bibitem[\protect\citeauthoryear{Rogers et al.}     {1996}]{rogers96}          Rogers, F. J.,Swenson, Fritz J., Iglesias, Carlos A., 1996, ApJ 456, 902
                                           
\bibitem[\protect\citeauthoryear{Rogers \& Nayfonov}     {2002}]{rogers02}          Rogers, F. J., Nayfonov, A., ApJ 576, 1064
                                           
\bibitem[\protect\citeauthoryear{Rosenthal et al.}  {1999}]{rosenthal99}       Rosenthal, C. S., Christensen-Dalsgaard, J., Nordlund, A., Stein, R. F., Trampedach, R., 1999, A\&A, 351, 689
                                           
\bibitem[\protect\citeauthoryear{Samadi et al.}        {2006}]{samadi06}             Samadi, R., Kupka, F., Goupil, M. J., Lebreton, Y., van't Veer-Menneret, C., 2006, A\&A, 445, 233
                                           
\bibitem[\protect\citeauthoryear{Schou}      {1998}]{schou99}           Schou, J., 1999, ApJ, 523, 181
                                           
\bibitem[\protect\citeauthoryear{Skartlien}      {2000}]{skart00}           Skartlien, R., ApJ 536, 465
                                           
\bibitem[\protect\citeauthoryear{Stein \& Nordlund}      {1998}]{stein98}           Stein, R.F., Nordlund, A., 1998, ApJ, 499, 914
                                           
\bibitem[\protect\citeauthoryear{Stein \& Nordlund}      {2006}]{stein06}           Stein, R.F., Nordlund, A., 2006, ApJ, 642, 1246
             
\bibitem[\protect\citeauthoryear{Stein et al.}      {2011}]{stein11}           Stein, R.F., Lagerfjard, A., Nordlund, A., Georgobiani, D., 2011, Sol. Phys., 268, 271

\bibitem[\protect\citeauthoryear{Stein \& Nordlund}      {2012}]{stein12}           Stein, R.F., Nordlund, A., 2012, ApJ, 753, 13

                              
\bibitem[\protect\citeauthoryear{Trampedach et al.} {2006}]{trampedach06}      Trampedach, R., Däppen, W., Baturin, V. A., 2006, ApJ 646, 560.
                                           
\bibitem[\protect\citeauthoryear{Trampedach et al.} {2006}]{trampedach13}      Trampedach, R., Asplund, M., Collet, R., Nordlund, A., Stein, R. F., 2013, ApJ 769, 18
                                           
\bibitem[\protect\citeauthoryear{Turck-Chieze et al.}         {1997}]{TC97}              Turck-Chi\`eze, S., Basu, S., Brun, A. S., Christensen-Dalsgaard, J., Eff-Darwich, A., Lopes, I., Pérez Hernández, F., Berthomieu, G., Provost, J., Ulrich, R. K., and 9 coauthors, 1997, Sol. Phys. 175, 247	


\end{thebibliography}
\end{document}